\documentclass[pre,twocolumn,preprintnumbers,amsmath,amssymb]{revtex4}

\usepackage{amsmath,latexsym,epsf,epsfig,graphicx}

\newcommand{\prt}{\partial}
\newcommand{\II}{\mbox{${\mathbb I}$}}

\def\UU{\mathbb U}
\def\S{\mathbb S}

\def\KK{\mathbb K}
\def\J{\mathbb J}

\newcommand{\mB}{{\mathbb B}}
\newcommand{\mA}{{\mathbb A}}

\newcommand{\D}{{\mathbb D}}

\newcommand{\rd}{{\rm d}}

\newcommand{\K}{{\mathcal K}}
\newcommand{\cP}{{\cal P}}
\newcommand{\dQ}{{\dot Q}}
\newcommand{\dS}{{\dot S}}
\newcommand{\dW}{{\dot W}}
\newcommand{\W}{{\mathcal W}}

\newcommand{\wt}{\hat{t}}

\def\A{\mathcal A}

\def\M{\mathcal M}

\def\der{\partial }

\def\ri{{\rm i}}

\def\prt{{\partial}}
\def\tr{{\rm Tr}}

\def\e{{\rm e}}

\def\rlb{\rangle_{{}_{\rm LB}}}

\begin{document}
\title{Microscopic Features of Bosonic Quantum Transport and Entropy Production} 

\author{Mihail Mintchev}
\affiliation{
Istituto Nazionale di Fisica Nucleare and Dipartimento di Fisica dell'Universit\`a di Pisa,\\
Largo Pontecorvo 3, 56127 Pisa, Italy} 

\author{Luca Santoni}
\affiliation{Institute for Theoretical Physics and Center for Extreme Matter and Emergent Phenomena, 
Utrecht University, Leuvenlaan 4, 3584 CE Utrecht, Netherlands}

\author{Paul Sorba} 
\affiliation 
{LAPTh, Laboratoire d'Annecy-le-Vieux de Physique Th\'eorique, 
CNRS, Universit\'e de Savoie,   
BP 110, 74941 Annecy-le-Vieux Cedex, France}
\bigskip 


\begin{abstract}

We investigate the microscopic features of bosonic quantum transport in a non-equilibrium steady state, 
which breaks time reversal invariance spontaneously. The analysis is based on the probability distributions, 
generated by the correlation functions of the particle current and the entropy 
production operator. The general approach is applied to an exactly solvable model with a point-like 
interaction driving the system away from equilibrium. The quantum fluctuations of the particle current 
and the entropy production are explicitly evaluated in the zero frequency limit. It is shown that all moments of the entropy 
production distribution are non-negative, which provides a microscopic version of the second law 
of thermodynamics. On this basis a concept of efficiency, taking into account all quantum fluctuations, 
is proposed and analysed. The role of the quantum statistics in this context is also discussed. 

\end{abstract}

\maketitle

\section{Introduction}

\begin{figure}[h]
\begin{center}
\begin{picture}(600,20)(155,360) 
\includegraphics[scale=0.92]{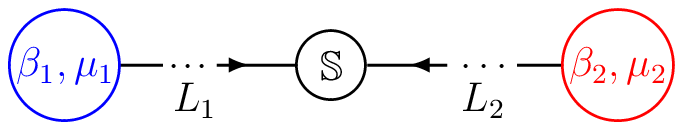}
\end{picture} 
\end{center}
\caption{(Color online) Two-terminal junction with bosonic heath baths connected with 
one-dimensional traps and a contact defect.} 
\label{fig1}
\end{figure}

This paper focuses on the basic microscopic properties of the particle and heat transport 
\and the relative entropy production in non-equilibrium bosonic quantum systems of the type shown 
in Fig. \ref{fig1}. The bulk of the system consists of two semi-infinite leads $L_i$, which are attached 
at infinity to two heat reservoirs (baths) $R_i$. The latter are both sources and sinks of particles 
and have large enough capacities, so that the particle emission and absorption 
do not change the (inverse) temperature $\beta_i\geq 0$ and the chemical potential 
$\mu_i$ of $R_i$. The contact between the two leads at $x=0$ represents 
a point-like impurity described by a unitary scattering matrix $\S$. 

The bosonic junction, shown schematically in Fig. \ref{fig1}, can be engineered by using  
ultracold Bose gases \cite{BDZ-08}-\cite{CS-16}, which attract recently much experimental 
and theoretical attention. The remarkable control over the interactions and the geometry of 
the samples in such experiments, as well as 
the absence of uncontrolled disorder, allow to explore unique aspects of many-body quantum physics. 
The advance in this rapidly developing area opens new horizons, including the possibility to create \cite{Z-04}-\cite{P-10} 
bosonic analogues of the conventional mesoscopic electronic devises like diodes and transistors (atomtronics). 

Coming back to the system in Fig. \ref{fig1}, one can imagine that the reservoirs 
$R_i$ contain ultracold atoms and are connected by two one-dimensional traps, 
which are implemented by confining electromagnetic 
fields and model the leads. The contact point between the two traps realises the impurity 
represented at the theoretical level by the scattering matrix $\S$. If the associated transmission 
probability $|\S_{12}|^2$ does not vanish, the system is away from equilibrium 
provided that the temperatures and/or chemical potentials of the two heat 
baths are different. 

The departure from equilibrium gives origin of incoming and 
outgoing matter and energy flows from the reservoirs $R_i$. Some decades ago 
Landauer \cite{L-57} and later B\"uttiker \cite{B-86} proposed an efficient method for studying 
these flows. The Landauer-B\"uttiker (LB) approach is based on the scattering matrix 
$\S$ and goes beyond the linear response approximation, thus 
representing an essential tool of modern quantum transport theory. The LB framework has been 
further generalised in \cite{A-80}-\cite{SI-86} and finds nowadays various applications, 
ranging from the computation 
of the noise power \cite{ML-92}-\cite{L-89} to the full counting statistics \cite{LL-92}-\cite{MSS-16}. 
Most of the quoted studies have been performed for fermionic systems. Triggered by the 
growing experimental activity with ultracold Bose cases, the investigation below is devoted to 
the bosonic case. In the next section we propose a general and universal approach to 
quantum transport at the microscopic level. In sections 3-5 we illustrate this approach at 
work, studying in detail an exactly solvable model. The role of the statistics is discussed 
in section 6. Finally, section 7 collects our conclusions and ideas for future investigations in the subject.

\section{General framework and strategy}

The basic observables, which characterise the quantum transport in the junction, are the particle current 
$j(t,x,i)$ flowing in the lead $L_i$ and the entropy production $\dS(t,x)$ in the whole system. They 
provide {\it local} and {\it global} information respectively, concerning the transport and its irreversibility. 
This information is codified in the correlation functions 
\begin{equation} 
w_n[j_i](t_1,x_1,...,t_n,x_n) = \langle j(t_1,x_1,i) \cdots j(t_n,x_n,i)  \rlb  \, , 
\label{cf1} 
\end{equation}
and 
\begin{equation}
w_n[\dS](t_1,x_1,...,t_n,x_n) = \langle \dS(t_1,x_1) \cdots \dS(t_n,x_n)  \rlb  \, , 
\label{cf2}
\end{equation} 
where the expectation value $\langle \cdots  \rlb$ is computed in the LB state \cite{M-11}. 
Following the standard approach \cite{K-87}-\cite{MSS-16} to full counting 
statistics, it is instructive to investigate the {\it zero frequency} limits $\W_n[j_i]$ of $w_n[j_i]$
and $\W_n[\dS]$ of $w_n[\dS]$, integrating the quantum fluctuations over long period 
of time. In this limit the dependence on the $2n$ space-time variables in 
(\ref{cf1},\ref{cf2}) drops out and one arrives at the following integral representations 
\begin{eqnarray} 
\W_n[j_i] &=& \int_0^\infty \frac{\rd \omega}{2\pi}\, \M_n[j_i] (\omega)\, , 
\label{cint} \\
\W_n[\dS] &=& \int_0^\infty \frac{\rd \omega}{2\pi}\, \M_n[\dS] (\omega)\, .
\label{eint}
\end{eqnarray}  
Here $\omega$ is the energy and $\M_n[j_i]$ and $\M_n[\dS]$ are the {\it moments} 
of two {\it probability distributions} $\varrho[j_i](\omega)$ and $\varrho[\dS](\omega )$, which govern 
the quantum fluctuations of the particle current and the entropy production respectively. 
The derivation of these distributions is a fundamental point of our approach. 
In fact, it turns out that $\varrho[j_i]$ and $\varrho[\dS]$ control respectively the 
elementary processes of emission and absorption of particles from the reservoirs 
and the associated entropy production. More precisely,  $\varrho[j_i]$ provides the 
quantum probabilities $\{p_k(\omega )\, :\, k=0,\pm 1,\pm 2,...\}$ for the aforementioned processes, 
whereas $\varrho[\dS]$ gives the values $\{\sigma_k(\omega )\, :\, k=0,\pm 1,\pm 2,...\}$ of 
the associated entropy production. 

In order to illustrate the above concepts at work, we propose and analyse in the paper an exactly 
solvable model. In this case we derive $p_k(\omega)$ and $\sigma_k(\omega )$ 
and show that they fully characterise the quantum transport and its efficiency at the 
{\it microscopic level}. A fundamental achievement of the paper 
is the explicit form of the probabilities $p_k(\omega)$ in terms of the Bose 
distributions $d_i(\omega)$ of the heat reservoirs $R_i$. 

The results about $\varrho[j_i]$ and $\varrho[\dS]$ shed new light on various central 
aspects of non-equilibrium quantum systems. First of all, they clarify the deep role of the statistics. 
The analysis in \cite{MSS-16} demonstrated that for fermions 
the Pauli exclusion principle implies $p_k(\omega )=0$ for all $k$ different from $0$ and $\pm 1$. 
We show below that this is not the case for bosons, where  
$p_k(\omega )\not= 0$ for all $k=0,\pm 1,\pm 2,...$. This feature is the 
microscopic origin of the different quantum transport properties of fermionic and 
bosonic systems.  
  
Another key aspect of our investigation concerns a remarkable feature of 
the entropy production distribution $\varrho[\dS]$. We prove below that 
all moments of this distribution are nonnegative, 
\begin{equation}
\M_n[\dS] (\omega) \geq 0 \, , \qquad \forall \; n=1,2,...
\label{slaw}
\end{equation}
The bound (\ref{slaw}) extends to the bosonic case our previous result \cite{MSS-17} for fermions 
and can be interpreted as a quantum counterpart of the second law of thermodynamics 
for the non-equilibrium bosonic system in Fig. \ref{fig1}. On this ground we propose an 
analog $\varepsilon_{II}$ of the concept of second law efficiency \cite{Bejan} from macroscopic thermodynamics.  
The knowledge of the distribution $\varrho[\dS]$ allows to separate at the 
fundamental level the processes with positive and negative entropy production and 
to extract from this information the coefficient $\varepsilon_{II}$, which takes into account 
the quantum fluctuations and characterises in an intrinsic way the transport in the system. 

In this paper we consider systems where the particle number and the total energy 
are conserved. These symmetries imply the existence of a conserved 
particle current $j(t,x,i)$ and energy current $\vartheta(t,x,i)$. 
The heat current is the linear combination 
\begin{equation}
q(t,x,i) = \vartheta (t,x,i) - \mu_i j(t,x,i)\, .  
\label{q}
\end{equation} 
Under this very general assumption about the symmetry content, one can prove \cite{MSS-14} that 
the junction in Fig. \ref{fig1} operates as {\it energy converter}. To be more explicit, 
let us consider the operator 
\begin{equation} 
\dQ = - \sum_{i=1}^2 q(t,0,i) \, , 
\label{qdot}
\end{equation} 
and let $\Phi$ be any state of the system. Then, if $\langle \dQ\rangle_\Phi <0$ the 
junction transforms heat to chemical energy. The opposite 
process takes place if instead $\langle \dQ\rangle_\Phi >0$. For a detailed study of 
this phenomenon of energy transmutation we refer to \cite{MSS-14}. 

An essential role in the general setup is played by the time reversal transformation 
\begin{equation}
T j(t,x,i) T^{-1} = -j(-t,x,i)\, , 
\label{tr1}
\end{equation}
where $T$ is an {\it anti-unitary} operator. We will show below that in the LB representation 
\begin{equation}
 \langle j(t,x,i) \rlb \not= - \langle j(-t,x,i) \rlb \, , 
\label{tr2}
\end{equation}
which implies that the LB state $\Omega_{{}_{\rm LB}}$ is not invariant 
under time reversal, $T\Omega_{{}_{\rm LB}} \not=\Omega_{{}_{\rm LB}}$. 
Consequently, the time reversal symmetry is spontaneously broken in 
the LB representation. The quantum transport process in the system is 
therefore irreversible, which gives rise to nontrivial entropy production 
described by the operator \cite{Callen, JP-01, N-07}
\begin{equation}
\dS (t,x) = - \sum_{i=1}^2 \beta_i\, q (t,x,i) \, .  
\label{dS}
\end{equation} 
It is worth mentioning that the currents depend on the 
lead $L_i$ where they are flowing, thus providing {\it local} information. The entropy production operator 
concerns instead the {\it global} system. Accordingly, the correlation functions (\ref{cf1}) 
refer to a single lead, whereas (\ref{cf2}) take into account all the {\it interference effects} 
between the heat currents in the two different leads 
$L_1$ and $L_2$. 

\section{Exactly solvable system} 

\subsection{The model}

The above considerations have a very general validity. In order to obtain concrete 
results however, one should fix the dynamics. 
In choosing among various possibilities, our guiding principle will be to focus on  
an exactly solvable model, where the zero frequency correlation functions $\W_n[j_i]$ 
and $\W_n[\dS]$ can be derived in explicit form for all $n$. For this purpose we 
consider the bosonic Schr\"odinger junction with a point-like defect. This system has 
already shown \cite{MSS-14}-\cite{MSS-16} to be a remarkable laboratory for testing general ideas about 
quantum transport. The dynamics along the {\it oriented} leads $L_i$ is fixed by 
the Schr\"odinger equation (the natural units $\hbar =c=k_{\rm B}=1$ are adopted 
throughout the paper)  
\begin{equation}
\left (\ri \prt_t +\frac{1}{2m} \prt_x^2\right )\psi (t,x,i) = 0\, ,  \qquad 
x < 0,\; \;  i=1,2\, , 
\label{eqm1}
\end{equation} 
and the canonical commutator 
\begin{equation}
[\psi(t,x_1,i_1)\, ,\, \psi^*(t,x_2,i_2)] = \delta_{i_1i_2}\, \delta(x_1-x_2)\, , 
\label{ccr1}
\end{equation}
where $*$ stands for Hermitian conjugation. The defect at $x=0$, which 
generates the interaction driving the system out of equilibrium, is fixed by the 
boundary condition 
\begin{equation} 
\lim_{x\to 0^-}\sum_{j=1}^2 \left [\lambda (\II-\UU)_{ij} +\ri (\II+\UU)_{ij}\prt_x \right ] \psi (t,x,j) = 0\, , 
\label{bc1} 
\end{equation} 
where $\II$ is the identity matrix, 
$\UU$ is a generic $2\times 2$ unitary matrix and $\lambda >0$ is a 
parameter with dimension of mass. Eq. (\ref{bc1}) defines the most general contact 
interaction between the two leads, which ensures \cite{ks-00} -\cite{h-00} unitary time evolution 
(self-adjointness of the bulk Hamiltonian). The associated scattering matrix is \cite{ks-00} -\cite{h-00}
\begin{equation} 
\S(k) = 
-\frac{[\lambda (\II - \UU) - k(\II+\UU )]}{[\lambda (\II - \UU) + k(\II+\UU)]} \, ,   
\label{S1}
\end{equation} 
$k$ being the particle momentum. More explicitly, 
\begin{eqnarray} 
\S(k) = \qquad \qquad \qquad \qquad \qquad 
\nonumber \\
\left(\begin{array}{cc}\frac{k^2 + \ri k (\eta_1-\eta_2)\cos(\vartheta)+\eta_1 \eta_2}{(k-\ri \eta_1)(k-\ri \eta_2)} 
& \frac{-\ri \e^{\ri \varphi} k (\eta_1-\eta_2)\sin(\vartheta)}{(k-\ri \eta_1)(k-\ri \eta_2)}\\ 
\frac{-\ri \e^{-\ri \varphi} k (\eta_1-\eta_2)\sin(\vartheta)}{(k-\ri \eta_1)(k-\ri \eta_2)} 
& \frac{k^2 - \ri k (\eta_1-\eta_2)\cos(\vartheta)+\eta_1 \eta_2}{(k-\ri \eta_1)(k-\ri \eta_2)}  \\ \end{array} \right)\, ,   
\nonumber \\
\label{NN0}
\end{eqnarray}
where $\varphi$ and $\vartheta$ are arbitrary angles and 
\begin{equation} 
\eta_i = \lambda \tan (\alpha_i)\, , 
\label{d4}
\end{equation} 
$\left (\e^{2\ri \alpha_1}, \e^{2\ri \alpha_2}\right )$ being the eigenvalues of 
$\UU$. The boundary bound states are the poles of (\ref{NN0}) located 
in the upper half-plane. We deduce from (\ref{d4}) that there are at most two 
bound states. The energy is bounded from below by 
\begin{equation}
\omega_{\rm min} = {\rm min}\left \{0, -\theta(\eta_1)\frac{\eta_1^2}{2m}, -\theta(\eta_2)\frac{\eta_2^2}{2m} \right \}\, , 
\label{spectrum}
\end{equation}
where $\theta$ is the Heaviside step function. 

In absence of bound states, the general solution of (\ref{eqm1}-\ref{bc1}) 
involves only the scattering component 
\begin{equation} 
\psi (t,x,i)  = \sum_{j=1}^2 \int_{0}^{\infty} \frac{dk}{2\pi } 
\e^{-\ri \omega (k)t}\, \Psi_{ij}(k;x) a_j (k) \, , 
\label{psi1} 
\end{equation} 
where $\omega(k) = \frac {k^2}{2m}$ is the dispersion relation, 
\begin{equation}
\Psi(k;x)=\left [ \e^{-\ri k x}\, \II +\e^{\ri k x}\, \S^*(k)\right ]\, , \quad k\geq 0\, , 
\label{ss}
\end{equation} 
and the operators $\{a_i(k),\, a^*_i(k)\, :\, k\geq 0,\, i=1,2\}$ generate a standard 
canonical commutation relation algebra $\A$. If bound states are present, 
the solution (\ref{psi1}) involves an additional term 
established in \cite{MSS-17an}. As explained there, 
this term contributes to the correlation functions (\ref{cf1},\ref{cf2}), but not to their 
zero frequency limits (\ref{cint},\ref{eint}) we are focusing on in this paper. For this 
reason a potential bound state contribution in (\ref{psi1}) can be safely neglected below.

\subsection{Basic observables}

Equations (\ref{eqm1}-\ref{bc1}) are invariant under $U(1)$-phase transformations and time translations, 
which imply particle number and total energy conservation. The associated conserved 
currents are 
\begin{equation}
j(t,x,i)= \frac{\ri }{2m} \left [ \psi^* (\partial_x\psi ) - 
(\partial_x\psi^*)\psi \right ]  (t,x,i)\, ,   
\label{curr}
\end{equation} 
and 
\begin{eqnarray}
\vartheta (t,x,i) = \frac{1}{4m} [\left (\partial_t \psi^* \right )\left (\partial_x \psi \right ) 
+ \left (\partial_x \psi^* \right )\left (\partial_t \psi \right ) 
\nonumber \\ - 
\left (\partial_t \partial_x \psi^* \right ) \psi - 
\psi^*\left (\partial_t \partial_x \psi \right ) ](t,x,i)\, ,  
\label{en1} 
\end{eqnarray} 
respectively. In order to derive the correlation functions (\ref{cf1},\ref{cf2}), one should 
express $j_x(t,x,i)$ and $\dS(t,x)$ in terms of the generators $\{a_i(k),\, a^*_i(k)\}$ 
of the algebra $\A$. Plugging the solution (\ref{psi1}) in (\ref{curr},\ref{dS}) one obtains 
\begin{eqnarray}
j(t,x,i)= \frac{\ri}{2m} \int_0^\infty \frac{\rd k}{2\pi} \int_0^\infty \frac{\rd p}{2\pi}\,  
\e^{\ri t [\omega(k) - \omega(p)]} 
\nonumber \\ 
\sum_{l,m=1}^2 a^*_l(k) \mA^i_{lm}(k,p,x) a_m(p) \, .
\label{curr1}
\end{eqnarray} 
\begin{eqnarray}
\dS(t,x)= \frac{\ri}{4m} \int_0^\infty \frac{\rd k}{2\pi} \int_0^\infty \frac{\rd p}{2\pi}\,  
\e^{\ri t [\omega(k) - \omega(p)]} \qquad \qquad \quad 
\nonumber \\ 
\sum_{i,l,m=1}^2 \beta_i[2\mu_i-\omega(k)-\omega(p)] a^*_l(k) \mA^i_{lm}(k,p,x) a_m(p)\, , 
\nonumber \\
\label{ds1}
\end{eqnarray} 
with 
\begin{eqnarray}
\mA^i_{lm}(k,p,x) \equiv \qquad \qquad \qquad  \qquad 
\nonumber \\
\Psi^*_{li}(k;x) \left [\der_{x} \Psi_{i m}\right ](p;x) - 
\left [\der_{x} \Psi^*_{li}\right ](k;x) \Psi_{im}(p;x) \, . 
\nonumber \\
\label{A}
\end{eqnarray}

The next step towards the derivation of the correlation functions (\ref{cf1},\ref{cf2}) 
is to fix a representation of the algebra $\A$.

\subsection{Correlation functions in the LB representation} 

Studying the physical setup in Fig. \ref{fig1} in a quantum mechanical context, 
Landauer \cite{L-57} and B\"uttiker \cite{B-86} suggested a non-equilibrium generalisation of the 
Gibbs representation of $\A$ (see e.g. \cite{M-11}) to systems, which exchange particles 
and energy with more then one heat reservoir. 
In what follows we call this generalisation the LB representation and refer to \cite{M-11} for 
a field theoretical construction of the associated Hilbert space. For deriving the 
expectation values of (\ref{curr1},\ref{ds1}) in the LB representation it is enough 
to compute the $2n$-point function 
\begin{equation}
\langle a^*_{l_1}(k_1) a_{m_1}(p_1)\cdots a^*_{l_n}(k_n) a_{m_n}(p_n)\rlb \, .
\label{cf3}
\end{equation}
Let us introduce for this purpose the $n\times n$ matrix 
\begin{equation} 
\mB_{ij} = 
\begin{cases} 
2\pi \delta (k_i-p_j)\delta_{l_im_j} d_{l_i}[\omega(k_i)]\, ,\qquad \qquad \; \; \; i\leq j\, , \\
2\pi \delta (k_i-p_j)\delta_{l_im_j}\left (1+ d_{l_i}[\omega(k_i)]\right )\, ,\qquad i > j\, ,\\ 
\end{cases} 
\label{B}
\end{equation} 
where 
\begin{equation}
d_l(\omega ) = \frac{1}{\e^{\beta_l (\omega -\mu_l)}-1}   
\label{bo}
\end{equation}  
is the familiar Bose distribution. Now, using the algebraic construction 
of the LB representation in \cite{M-11}, one can show that the correlation function 
(\ref{cf3}) is the {\it permanent} of the matrix $\mB$,  
\begin{equation}
\langle a^*_{l_1}(k_1) a_{m_1}(p_1)\cdots a^*_{l_n}(k_n) a_{m_n}(p_n)\rlb = {\rm perm}[\mB] \, . 
\label{cf7}
\end{equation}
It is perhaps useful to recall the explicit form 
\begin{equation}
{\rm perm}[\mB]= \sum_{\sigma_i \in \cP_n} \prod_{i=1}^n \mB_{i \sigma_i} \, , 
\label{perm}
\end{equation}
where $\cP_n$ is the set of all permutations of $n$ elements. 

By means of (\ref{cf7},\ref{perm}) one easily derives the one-point current correlation function 
\begin{equation}
\langle j(t,x,i) \rlb = 
\int_0^\infty \frac{\rd \omega}{2\pi} \sum_{l=1}^2 \left (\delta_{il}-|\S_{il}(\sqrt {2m\omega})|^2\right ) d_l(\omega)\, . 
\label{tr3}
\end{equation}
The right hand side of (\ref{tr3}) implies (\ref{tr2}), 
establishing a basic characteristic feature of the LB representation - the spontaneous 
breakdown of time reversal invariance. Let us observe in this respect that the dynamics (\ref{eqm1}) 
is time reversal invariant. The same holds for the boundary condition (\ref{bc1}), provided that 
besides being unitary the matrix $U$ is also symmetric. 

We would like to comment at this stage on the range of the chemical potentials $\mu_i$. 
In order to avoid singularities, indicating condensation phenomena, 
we assume that the particle density $d_l(\omega )$ in the reservoir $R_l$ 
is positive for all $\omega \geq 0$. Therefore $\mu_l <0$ in what follows.

\subsection{The zero frequency limit}

Since at this point the correlation functions (\ref{cf1},\ref{cf2}) can be treated in the same way, 
we introduce the notation $\{w_n[\zeta]\, :\, \zeta = j_i,\dS;\, n=1,2,...\}$. 
Time translation invariance implies that $w_n[\zeta]$ 
depend actually only on the time differences  
\begin{equation}
\wt_k \equiv t_k - t_{k+1}\, , \quad \, k=1,...,n-1\, .  
\label{td}
\end{equation}
which allows to introduce for $n\geq 2$ the frequency $\nu$ via the Fourier transform 
\begin{eqnarray} 
\W_n[\zeta](x_1,...,x_n;\nu ) = 
\int_{-\infty}^{\infty} \rd \wt_1 \cdots   \int_{-\infty}^{\infty} \rd \wt_{n-1} 
\nonumber \\
\e^{-\ri \nu (\wt_1+\cdots \wt_{n-1})} w_n[\zeta](t_1, x_1,...,t_n,x_n)\, .  
\label{cf8}
\end{eqnarray} 
Following the classical studies \cite{ML-92}-\cite{MSS-15} of the fermionic quantum noise, 
which have been extended in \cite{GGM-03} to the current cumulants with $n>2$ 
and applied in the framework of full counting statistics \cite{K-87}-\cite{MSS-16}, 
we perform the zero-frequency limit 
\begin{equation}
\W_n[\zeta] = \lim_{\nu \to 0^+} \W_n[\zeta](x_1,...,x_n;\nu ) \, .
\label{cf9}
\end{equation} 
In the limit (\ref{cf9}) the quantum fluctuations are integrated over 
the whole time axes and it turns out that the position dependence drops out. 
Because of this relevant simplification, the zero frequency regime is 
intensively explored also in experiments. 

The derivation of $\W_n[j_i]$ and $\W_n[\dS]$ in explicit form is straightforward but long. 
For this reason we summarise below only the main steps of the procedure: 

(i) using (\ref{curr1},\ref{ds1},\ref{A}) and (\ref{cf7}) one first obtains a representation of the correlation 
function $w_n[\zeta](t_1, x_1,...,t_n,x_n)$, involving $n$ integrations over $k_i$ and 
$n$ integrations over $p_j$;

(ii) by means of the delta functions in (\ref{B}) one eliminates all $n$ integrals in $p_j$; 

(iii) plugging the obtained expression in (\ref{cf8}), one performs all $(n-1)$ integrals in $\wt_l$; 

(iv)  at $\nu=0$ the latter produce $(n-1)$ delta-functions, which allow to eliminate all the integrals 
over $k_i$ except one, for instance that over $k_1=k$;   

(v) since now the matrix (\ref{A}) must be evaluated at $k=p$, the $x$-dependence drops out and 
one finds  
\begin{equation}
\mA^i_{lm}(k,k,x) = 
-2 k [\delta_{li}\delta_{ij} - \S_{li}(k) {\overline \S}_{ji}(k)]\, , 
\label{A1}
\end{equation} 
the bar indicating complex conjugation;

(vi) switch to the variable $\omega =k^2/2m$ in the integral over $k$; 

(vii) introduce the $2\times 2$ matrix  
\begin{eqnarray}
\J_{11}(\omega )&=&-\J_{22}(\omega ) = |\S_{12}(\sqrt {2 m \omega})|^2 \equiv \tau (\omega )\, , 
\label{T1} \\
\J_{12}(\omega )&=&{\overline \J}_{21}(\omega )=-\S_{11}(\sqrt {2 m \omega}) {\overline\S}_{12}(\sqrt {2 m \omega})\, , 
\label{T2}
\end{eqnarray} 
where $\tau(\omega)$ is the {\it transmission probability}; 

(viii) introduce finally the $n\times n$ matrix 
\begin{equation} 
\D_{ij} (\omega ;l_1,...,l_n)= 
\begin{cases} 
\J_{l_jl_i}(\omega ) d_{l_j}(\omega )\, ,\qquad \qquad \qquad \; i\leq j\, , \\
\J_{l_jl_i}(\omega ) d_{l_j}(\omega )\left [1+ d_{l_i}(\omega )\right ]\, ,\quad i > j\, ,\\ 
\end{cases} 
\label{D}
\end{equation} 
and define the sum of permanents 
\begin{equation}
\K_n(\omega ) =  \sum_{l_1,...,l_n=1}^2 {\rm perm}[\D (\omega ;l_1,...,l_n)]\, . 
\label{K}
\end{equation} 
Using the matrix (\ref{D}) and the definition (\ref{perm}) of permanent, one can explicitly derive 
each term of the sequence $\{\K_n(\omega )\, :\, n=1,2,...\}$, which 
will play an important role in what follows. The first few terms are: 
\begin{eqnarray} 
\K_1(\omega) &=& \tau(\omega) c_1(\omega)\, , 
\nonumber \\
\K_2(\omega) &=& \tau(\omega)\left [c_2(\omega) +2 c_1^2(\omega)\tau(\omega)\right ], ,  
\nonumber \\
\K_3(\omega) &=& \tau^2(\omega)c_1(\omega)\left [1+6c_2(\omega) +6 c_1^2(\omega)\tau(\omega)\right ]\, , 
\nonumber \\
\label{K1}
\end{eqnarray} 
where $c_i$ are the frequently used below combinations 
\begin{eqnarray}
c_1(\omega) &=& d_1(\omega)-d_2(\omega)\, ,
\label{c1}\\
c_2(\omega) &=& d_1(\omega)+d_2(\omega)+2d_1(\omega) d_2(\omega)\, . 
\label{c2}
\end{eqnarray}

Performing the above eight steps, one obtains for the particle current the integral representation  
(\ref{cint}) with 
\begin{eqnarray} 
\M_n[j_1] &=& \K_n (\omega )\, , 
\label{mc1} \\ 
\M_n[j_2] &=& (-1)^n \K_n(\omega )\, ,
\label{mc2} 
\end{eqnarray}
and for the entropy production (\ref{eint}) with
\begin{equation} 
\M_n[\dS] = [\gamma_{21}(\omega )]^n \K_n(\omega ) \, ,
\label{me1} 
\end{equation}
where 
\begin{equation}
\gamma_{ij}(\omega ) = (\beta_i-\beta_j)\omega - (\beta_i\mu_i-\beta_j\mu_j) \, . 
\label{gamma}
\end{equation}
As already observed in the introduction, $\M_n[j_i]$ and $\M_n[\dS]$ given by 
(\ref{mc1},\ref{mc2}) and (\ref{me1}), are the moments of the probability distributions 
$\varrho [j_i]$ and $\varrho[\dS]$. The goal of the next section is to reconstruct 
these distributions from the moments and uncover the physical information codified therein.

\section{Probability distributions} 

\subsection{Moment generating functions}

The general problem now is to find a function $\varrho[\zeta]$ such that 
\begin{equation} 
\M_n[\zeta ] = \int_{-\infty}^\infty \rd \sigma \, \sigma^n \varrho[\zeta] (\sigma)\, , \qquad n=0,1,... \, , 
\label{pr00}
\end{equation} 
where $\M_n[\zeta]$ are given for $n\geq 1$ by (\ref{mc1}-\ref{me1}) and 
\begin{equation}
\M_0[\zeta] = 1 
\label{pr0}  
\end{equation} 
is a normalisation condition. As well known \cite{ST-70}, $\varrho[\zeta]$ is given by the Fourier transform 
\begin{equation} 
\varrho[\zeta] (\sigma) = \int_{-\infty}^\infty \frac{\rd \lambda}{2\pi}\, \e^{-\ri \lambda \sigma}\, \chi[\zeta] (\lambda ) 
\label{pr1}
\end{equation} 
of the moment generating function 
\begin{equation} 
\chi[\zeta] (\lambda) = \sum_{n=0}^\infty \frac{(\ri \lambda)^n}{n!}\, \M_n[\zeta]\, . 
\label{pr2}
\end{equation} 
We will proceed therefore by determining first $\chi[\zeta]$ from the corresponding 
moments, given by (\ref{mc1}-\ref{me1}), and after that performing the Fourier transform (\ref{pr1}). 

{}For deriving the moment generating function $\chi[j_1](\lambda)$ in explicit form 
we apply to our case the technique, developed by Glauber in quantum optics for 
the counting statistics of photons (see. e.g. \cite{Glauber}). First we 
introduce two auxiliary bosonic oscillators $\{a^*_i,a_i\:, :\, i=1,2\}$, 
satisfying the commutation relations 
\begin{equation}
[a_i\, ,\, a^*_j]=\delta_{ij}\, , \qquad [a_i^*\, ,\, a^*_j] = [a_i\, ,\, a_j]=0\, . 
\label{Al1}
\end{equation} 
Using these oscillators one can generate the sum of permanents in the right 
hand side of (\ref{K}). In fact, setting  
\begin{equation}
K(\omega )= \sum_{i=1}^2 a^*_i\, \KK_{ij}(\omega ) a_i \, , \qquad 
J(\omega )= \sum_{i,j=1}^2 a^*_i\, \J_{ij}(\omega)\, a_j \, ,  
\label{A2}
\end{equation} 
where  
\begin{equation}
\KK_{ij}(\omega) = \beta_i(\omega-\mu_i) \delta_{ij}\, , 
\label{A3}
\end{equation}
and $\J(\omega )$ is given by (\ref{T1},\ref{T2}), one can verify by means of (\ref{Al1}) that 
\begin{equation}
\sum_{l_1,...,l_n=1}^2 {\rm perm}[\D (\omega ;l_1,...,l_n)] =   
\frac{\tr \left [\e^{-K(\omega )} J(\omega)^n \right ]}{\tr \left [\e^{-K(\omega)}\right ]}\, .
\label{A4}
\end{equation} 
Equations (\ref{K},\ref{mc1},\ref{pr2},\ref{A4}) now imply the following trace representation for the 
moment generating function  
\begin{equation}
\chi[j_1](\lambda ) = \frac {\tr \left [\e^{-K(\omega )} \e^{\ri \lambda J(\omega )}\right ]}{\tr \left [\e^{-K(\omega)}\right ]}\, . 
\label{A5}
\end{equation} 
Observing that 
\begin{equation}
[K(\omega )\, ,\, J(\omega)]= \sum_{i,j=1}^2 a^*_i \left ([\KK\, ,\, \J]\right )_{ij} a_j \, , 
\label{A6}
\end{equation}
one can apply the result of \cite{K-03} for traces of the type (\ref{A5}) and obtain the alternative determinant representation 
\begin{eqnarray} 
\chi[j_1](\lambda ) &=& \frac{\det \left [1-\e^{\KK(\omega )}\right ]}{\det \left [1-\e^{\KK(\omega )} \e^{\ri \lambda \J(\omega )}\right ]}
\nonumber \\
&=& \frac{1}{\det \left [1-\left (1-\e^{\KK(\omega )}\right )^{-1}\e^{\KK(\omega )} 
\left ( \e^{\ri \lambda \J(\omega )}-1\right )\right ]}\, , 
\nonumber \\
\label{A7}
\end{eqnarray} 
which is more manageable. Indeed, using that 
\begin{equation}
\left [\left (1-\e^{\KK(\omega )}\right )^{-1}\e^{\KK(\omega )} \right ]_{ij} = d_i(\omega )\delta_{ij}
\label{A8}
\end{equation}
and once again the explicit form (\ref{T1},\ref{T2}) of the matrix $\J(\omega )$, one finally gets 
\begin{equation} 
\chi[j_1] (\lambda) = \frac{1}{1-\ri c_1{\sqrt \tau} \sin(\lambda {\sqrt \tau}) -c_2 [\cos(\lambda {\sqrt \tau})-1]}\, . 
\label{pr3}
\end{equation} 
For conciseness we omit here and in what follows the dependence of $\tau$ and $c_i$ on 
the energy $\omega$. 

Analogously, for the entropy production generating function one finds 
\begin{equation} 
\chi[\dS] (\lambda) = \frac{1}{1-\ri c_1{\sqrt \tau} \sin(\lambda \gamma_{21} {\sqrt \tau}) -
c_2 [\cos(\lambda  \gamma_{21} {\sqrt \tau})-1]}\, .  
\label{epr1}
\end{equation} 

According to (\ref{pr1}) the probability distributions $\varrho[j_1]$ and $\varrho[\dS]$, we are looking for, 
are obtained by performing the Fourier transform of (\ref{pr3},\ref{epr1}).

\subsection{Particle current distribution}

Since the right hand side of (\ref{pr3}) is a periodic function with period $2\pi/\sqrt \tau$, 
the generating function has the Fourier expansion 
\begin{equation}
\chi[j_1] (\lambda)= \sum_{k=-\infty}^{\infty} p_k\, \e^{\ri k \lambda \sqrt \tau} \, . 
\label{pr4}
\end{equation}
The coefficients $\{p_{\pm n}\, :\, n=0,1,...\}$ can be deduced from (\ref{pr3}) and read 
\begin{equation}
p_{\pm n} = \frac{c_\pm^n}{1+c_2} \sum_{j=0}^{\infty}\binom{2j+n}{j} (c_+ c_-)^j \, , 
\label{pr5}
\end{equation} 
where 
\begin{equation}
c_\pm= \frac{(c_2\pm c_1\sqrt \tau)}{2(1+c_2)} \, . 
\label{pr6}
\end{equation} 
Using that $\mu_i<0$ and $0\leq \tau \leq 1$ one can show that 
\begin{equation}
c_2> 0\, , \qquad c_\pm >0\, , \qquad c_\pm <1\, , \qquad c_+c_- <1/4\, , 
\label{inequalities}
\end{equation} 
which imply that the series (\ref{pr5}) is convergent and  
$p_k > 0$ for all $k$. Indeed, in closed form one has the Gauss hypergeometric function 
\begin{equation}
p_{\pm n} = \frac{c_\pm^n}{1+c_2}\, {}_2F_1 \left [\frac{1+n}{2}, \frac{2+n}{2}, n+1,4c_+c_-\right ] > 0\, . 
\label{pr7}
\end{equation} 

From (\ref{pr4}) one deduces that the probability distribution $\varrho[j_1]$ is the {\it Dirac comb} function 
\begin{equation}
\varrho[j_1](\xi ) = \sum_{k=-\infty}^{\infty} p_k\, \delta(\xi -k\sqrt \tau) \, , 
\label{pr8}
\end{equation} 
where the weights $p_k$ are given by (\ref{pr7}) and satisfy 
\begin{equation} 
\sum_{k=-\infty}^{\infty} p_k = 1\, . 
\label{pr9}
\end{equation}
In fact, (\ref{pr3}) implies on one hand  
\begin{equation} 
\int_{-\infty}^{\infty} \rd \xi \varrho [j_1](\xi) = \chi[j_1] (0) = 1\, .
\label{pr10}
\end{equation} 
On the other hand, integrating (\ref{pr8}) one gets 
\begin{equation}
\int_{-\infty}^{\infty} \rd \xi \varrho[j_1] (\xi) = \sum_{k=-\infty}^{\infty} p_k \, ,  
\label{pr11}
\end{equation}
which completes the argument. Summarising, since in addition $p_k > 0$, the coefficients 
$p_k$ represent probabilities, whose physical meaning will be uncovered below. 

Finally, using (\ref{mc2}) one concludes that 
\begin{equation} 
\chi[j_2](\lambda) = \chi[j_1](-\lambda)\, , 
\label{cpr12}
\end{equation}
which implies in turn that 
\begin{equation} 
\varrho[j_2](\xi) = \varrho[j_1](-\xi)\, .  
\label{cpr13}
\end{equation}
Equivalently, $\varrho[j_1]$ and $\varrho[j_2]$ are related by 
\begin{equation}
p_k \mapsto p_{-k} \, . 
\label{cpr14}
\end{equation}

\subsection{Entropy production distribution}

\begin{figure}[ht]
\begin{center}
\includegraphics[scale=0.7]{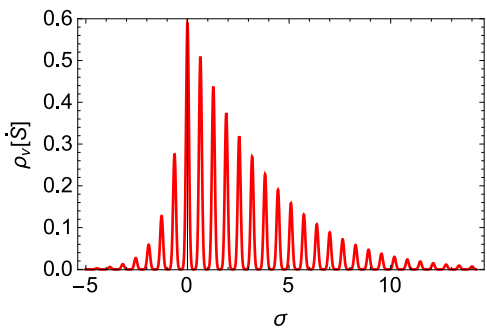}
\end{center}
\caption{(Color online) The smeared distribution $\varrho_\nu$ with $\nu=200$, $\gamma_1=1$, $\gamma_2=0.1$ and $\tau=1/2$.} 
\label{fig2}
\end{figure}

Employing (\ref{epr1}), a straightforward extension of the analysis in the previous subsection leads to 
\begin{equation}
\varrho[\dS](\sigma ) = \sum_{k=-\infty}^{\infty} p_k\, \delta(\sigma -k\, \gamma_{21}{\sqrt \tau}) \, , 
\label{epr2}
\end{equation} 
with the same probabilities $p_{\pm n}$ given by (\ref{pr7}). This property is the first 
indication that the probabilities $\{p_{\pm n}\, :\, n=0,1...\}$ carry universal and fundamental information 
about the quantum transport at the microscopic level. We postpone the detailed discussion 
of this issue to the next subsection, focussing here on the possibility to illustrate 
graphically the behavior of the distribution (\ref{epr2}). For this purpose it is 
convenient to introduce the $\delta$-sequence 
\begin{equation} 
\delta_\nu(\sigma ) = \frac{\nu}{\sqrt \pi}\, \e^{-\nu^2 \sigma^2}\, , \qquad \nu>0\, , 
\label{delta1}
\end{equation}
and the {\it smeared} distribution 
\begin{equation}
\varrho_\nu[\dS](\sigma ) = \sum_{k=-\infty}^{\infty} p_k\, \delta_\nu (\sigma -k\, \gamma_{21}{\sqrt \tau}) \, , 
\label{epr3}
\end{equation} 
As well known, for $\nu \to \infty$ one has $\varrho_\nu \to \varrho$ 
in the sense of generalised functions.  The plot of $\varrho_\nu$, reported in Fig. \ref{fig2}, 
nicely illustrates the physics discussed in the next subsection. 

\subsection{Physical interpretation of the probabilities $p_k$} 

It is instructive in this section to restore the electric charge $e$ in the current (\ref{curr}) 
by $j \mapsto e j$. The distribution (\ref{pr8}) now takes the form 
\begin{equation}
\varrho[j_1](\xi ) = \sum_{k=-\infty}^{\infty} p_k\, \delta(\xi -k e \sqrt \tau) \, , 
\label{pi1}
\end{equation} 
where the parameter $\xi$ measures the charge which is transferred between the two reservoirs. 
Without loss of generality we can assume that $\xi$ is positive if the particles are emitted from $R_2$ and 
absorbed by $R_1$ and is negative for the process in the opposite direction. The argument of the 
delta function in (\ref{pi1}) suggest a simple interplay between the transport on one hand and the
processes of emission and absorption on the other hand. Suppose that $R_2$ emits in the system the charge 
$ke$. Because of the defect, the part $ke\sqrt \tau$ is transmitted and absorbed by $R_1$. The rest 
$ke(1-\sqrt \tau)$ is reflected by the defect and reabsorbed by $R_2$. This is a purely quantum scattering effect. 
For $\tau=1$ the defect is fully transparent 
and the charge emitted from one reservoir is totally absorbed by the other one. Finally, the term $k=0$ in (\ref{pi1})
describes the emission and absorption of particles by the same reservoir, thus corresponding 
to a vanishing charge transfer $\xi=0$. 

Summarising, the probabilities $p_k$ fully characterise the elementary processes of 
particle emission and absorption by the heat reservoirs, which provide in turn 
the common basis for all types of transport in the junction. In fact, the probability 
distributions for the energy and heat currents $\vartheta_1$ and $q_1$ in the 
lead $L_1$ are obtained from (\ref{pi1}) by the substitution 
\begin{equation} 
e \longmapsto  
\begin{cases} 
\omega \, ,\qquad \qquad \qquad  \; \;  {\rm for}\quad  \varrho [\vartheta_1]\, , \\
\omega -e \mu_1\, ,\qquad \quad \; \; \; \, {\rm for}\quad  \varrho [q_1]\, ,\\ 
\end{cases} 
\label{pi2}
\end{equation}
which confirms the universal character of the the probabilities $p_k$.

\subsection{Microscopic quantum version of the second law} 

From (\ref{epr3}) one infers for the entropy production the values 
$\{\sigma_k=k\gamma_{21}\sqrt \tau\, :\, k=0,\pm 1,\pm2,...\}$. 
Suppose now that $\gamma_{21}>0$. Accordingly, we call $R_2$ the ``hot" 
reservoir and $R_1$ the ``cold" one. In this case we deduce from (\ref{epr2}) that $p_k$ with $k>0$ 
are associated with the transmission from the hot reservoir to the cold one, leading to 
positive entropy production $\sigma_k=k\gamma_{21}\sqrt \tau >0$. For $k<0$ instead, 
$p_k$ correspond to the transport from the cold to the hot reservoir, which generates negative 
entropy production $\sigma_k = k\gamma_{21}\sqrt \tau <0$. For $k=0$ there is no particle exchange 
between the two reservoirs and consistently the entropy production vanishes. 

Analysing $c_\pm$, given by (\ref{pr6}), it is easy to show that for $k>0$ 
\begin{equation}
\gamma_{21}>0 \quad \Longrightarrow \quad c_+>c_- \quad \Longrightarrow \quad p_{k} > p_{-k} \, . 
\label{prob3}
\end{equation}
The processes with positive entropy production thus dominate that with negative one. This feature 
is illustrated in Fig. \ref{fig2} and suggests that like in the fermionic case \cite{MSS-17} all moments 
(\ref{me1}) of the probability distribution (\ref{epr2}) are non-negative (\ref{slaw}). For proving 
this bound we denote by $\Delta(\lambda )$ the 
denominator of $\chi[\dS](\lambda )$, given by (\ref{epr1}), and observe that 
\begin{eqnarray}
\Delta(-\ri \lambda) &=& 1- c_1{\sqrt \tau} \sinh(\lambda \gamma_{21} {\sqrt \tau}) -
c_2 [\cosh(\lambda  \gamma_{21} {\sqrt \tau})-1] 
\nonumber \\
&=&\sum_{n=0}^\infty  \lambda^n a_n\, , 
\label{m1}
\end{eqnarray}
where 
\begin{equation} 
a_n= 
\begin{cases} 
1 \, ,\qquad \qquad \qquad \qquad \; \; \; n=0\, , \\
-\frac{\gamma_{21}^{2k-1} \tau^k}{(2k-1)!} c_1\, ,\qquad \quad \; \; \; \, n=2k-1\, ,\\ 
-\frac{\gamma_{21}^{2k} \tau^k}{(2k)!} c_2\, ,\qquad \qquad \; \; \; \;   n=2k\, . \\
\end{cases} 
\label{m2}
\end{equation}
On the other hand, 
\begin{equation}
\chi[\dS](-\ri \lambda )= \sum_{n=0}^\infty  \lambda^n b_n\, ,\qquad b_n=\frac{\M_n[\dS]}{n!} \, , 
\label{m3}
\end{equation} 
and using the identity 
\begin{equation}
\Delta(-\ri \lambda ) \chi[\dS](-\ri \lambda) = 1\, , 
\label{m4}
\end{equation} 
one gets for $n\geq 1$ the recursive relation 
\begin{equation}
b_n =  
-a_n-a_{n-1}b_1 - a_{n-2} b_2 - \cdots - a_1 b_{n-1}    
\label{m5}
\end{equation} 
with 
\begin{equation}
b_0=\M_0[\dS]=1 \, .  
\label{m5a}
\end{equation} 
At this point the inequality $\M_n[\dS]\geq 0$ follows from (\ref{m5}) by induction, 
observing that $\mu_i<0$ implies that 
\begin{equation}
\gamma_{21} c_1 \geq 0\, ,\quad c_2\geq 0 \quad  \Longrightarrow \quad 
a_n \leq 0 \, , \qquad \forall \; n \geq 1\, . 
\label{m6}
\end{equation} 
For the even moments $\M_{2k}[\dS]$ this feature is a direct consequence of  
the fact that (\ref{epr2}) is a probability distribution ($0<p_k<1$), 
but for the odd ones $\M_{2k-1}[\dS]$, this is not at all automatic 
and represents a characteristic feature of the distribution $\varrho[\dS]$, 
governing the entropy production fluctuations. Since $\M_1[\dS]$ is the mean 
value of the entropy production, the bound (\ref{slaw}) can be interpreted \cite{MSS-17} as 
a quantum counterpart of the second law of thermodynamics for a system 
in a fixed steady state, which implements the contact with two heat reservoirs as 
shown in Fig. \ref{fig1}. We would like to mention in this respect that a quantum 
version of the second law, relative 
to the transition between different states of a system in contact 
with one heat bath, has been proposed in \cite{BHNOW-15, CSHO-15}. In that case 
there is actually a whole family of ``second laws", each of them enforcing 
a specific physical constraint on the thermodynamic evolution. 

We conclude this section by an observation concerning the influence of a 
hypothetical {\it classical} measuring devise on the system under consideration. 
The simplest way to implement such a devise is to introduce  \cite{LLL-96}-\cite{ELB-09} 
in (\ref{eqm1}) a minimal coupling $\ri \der_x \longmapsto \ri \der_x + A(x)$ with a suitable 
{\it classical} external field $A(x)$. The study of this new setup can be performed 
following mutatis mutandis the above analysis and is beyond the scope of this paper, focussed 
exclusively on the quantum behavior. In the fermionic case the impact  
of a classical field $A(x) \sim \delta(x)$ on $\varrho[j_i]$ and $\varrho[\dS]$ has been discussed 
in detail in \cite{LC-03} and \cite{MSS-16} respectively.

\section{Efficiency} 

At the quantum level the question of efficiency has been addressed in the past mainly 
by studying the mean value $\langle q(t,x,i)\rlb $ of the heat currents in the two leas $L_i$. 
Unfortunately, $\langle q(t,x,i)\rlb $ does not keep trace of the quantum fluctuations, 
which are expected to affect the quantum efficiency and whose presence is actually 
the relevant novelty with respect to the classical case. 

Our main objective here is to introduce and study a suitable quantity, which describes 
the transport efficiency at the microscopic level and captures the quantum fluctuations. 
To this end we use the set of probabilities $\{p_k\, :\, k=0,\pm 1, \pm 2,...\}$ derived above and 
take advantage of the fact that our formalism provides 
the values of the positive and negative entropy productions $\sigma^{\rm tot}_\pm$ separately 
and not only the value of their sum $\sigma^{\rm tot}_+ +\sigma^{\rm tot}_-$. In fact, 
the probability distribution (\ref{epr2}) implies that for $\gamma_{21}>0$ 
the probabilities $p_{n\geq 1}$ and $p_{n\leq -1}$ 
correspond to positive and negative entropy production respectively. For $\gamma_{21}<0$ one has that  
$p_{n\geq 1}$ and $p_{n\leq -1}$ exchange their role. Therefore 
\begin{eqnarray}
\sigma_\pm (\omega ) = \qquad \qquad \qquad \qquad
\nonumber \\
\pm \theta (\gamma_{12}) \gamma_{12} \sqrt{\tau} \sum_{n=1}^\infty n\, p_{\mp n} 
\pm \theta (\gamma_{21}) \gamma_{21} \sqrt{\tau} \sum_{n=1}^\infty n\, p_{\pm n}\, , 
\label{s1} 
\end{eqnarray} 
give the positive and negative entropy production at energy $\omega$. Consequently, 
the total positive/negative entropy production in the system is 
\begin{equation}
\sigma^{\rm tot}_\pm = \int_0^\infty \rd \omega \, \sigma_\pm (\omega ) \, . 
\label{s3}
\end{equation}
Because of (\ref{prob3}) one has 
\begin{equation}
\sigma_+(\omega)>-\sigma_-(\omega )>0 \quad  \Longrightarrow \quad \sigma_+^{\rm tot}>-\sigma_-^{\rm tot}>0 \, . 
\label{s4}
\end{equation} 

The main idea now is to extend and adapt the concept of {\it second law} macroscopic efficiency 
(see e.g. \cite{Bejan}) for heat engines to our case. In order to recall briefly this concept, 
let us consider for a moment a classical 
heat engine in contact with two heat baths as shown in Fig. \ref{fig1}. Let us denote the work 
transfer rate of the engine by $\dW>0$. Moreover, let 
$\dW_{\rm rev}$ be the value of $\dW$ in the limit of reversible operation. 
Then, the second law efficiency is defined by \cite{Bejan} 
\begin{equation} 
\eta_{\rm II} = \frac{\dW}{\dW_{\rm rev}} \, . 
\label{he1}
\end{equation} 
It is perhaps useful to recall also that the more familiar {\it first law} efficiency $\eta_{\rm I}$ 
is given in terms of $\eta_{\rm II}$ by \cite{Bejan} 
\begin{equation} 
\eta_{\rm I} = \eta_{\rm II} (1-r)\, , \qquad r \equiv \frac{\beta_1}{\beta_2}\, , \qquad \beta_2 \geq \beta_1\, .  
\label{he2}
\end{equation} 
The second law of thermodynamics states that $\dW \leq \dW_{\rm rev}$, implying $\eta_{\rm II} \leq 1$.  
The value $\eta_{\rm II} = 1$ is reached in the limit of reversibility. This is the fundamental 
property we would like to preserve when introducing a concept of efficiency 
for the quantum transport in the junction in Fig. \ref{fig1}, 
where instead of the work transfer rates $\dW$ and $\dW_{\rm rev}$,  
we know the entropy productions $\sigma_\pm^{\rm tot}$. 
At this point the quantum second law in the form (\ref{s4}) suggest to consider the quantity  
\begin{equation} 
\varepsilon_{\rm II} = -\frac{\sigma_-^{\rm tot}}{\sigma_+^{\rm tot}} \, , 
\label{s5}
\end{equation} 
which satisfies $0 \leq \varepsilon_{\rm II} \leq 1$ and has the desired reversibility limit 
\begin{equation}
\sigma_+^{\rm tot} + \sigma_-^{\rm tot} = 0 \quad \Longrightarrow \quad \varepsilon_{\rm II}  =1\, . 
\label{s6}
\end{equation} 
In addition, setting 
\begin{equation} 
\varepsilon_{\rm I} = \varepsilon_{\rm II} (1-r) = -\frac{\sigma_-^{\rm tot}}{\sigma_+^{\rm tot}}(1- r)\, , 
\label{s7}
\end{equation} 
we conclude that, like in the case of heat engines, $\varepsilon_{\rm I}$ can not exceed 
the familiar Carnot efficiency $\eta_{\rm C}=1-r$, recovered in the regime (\ref{s6}) of reversibility. 

\begin{figure}[ht]
\begin{center}
\includegraphics[scale=0.455]{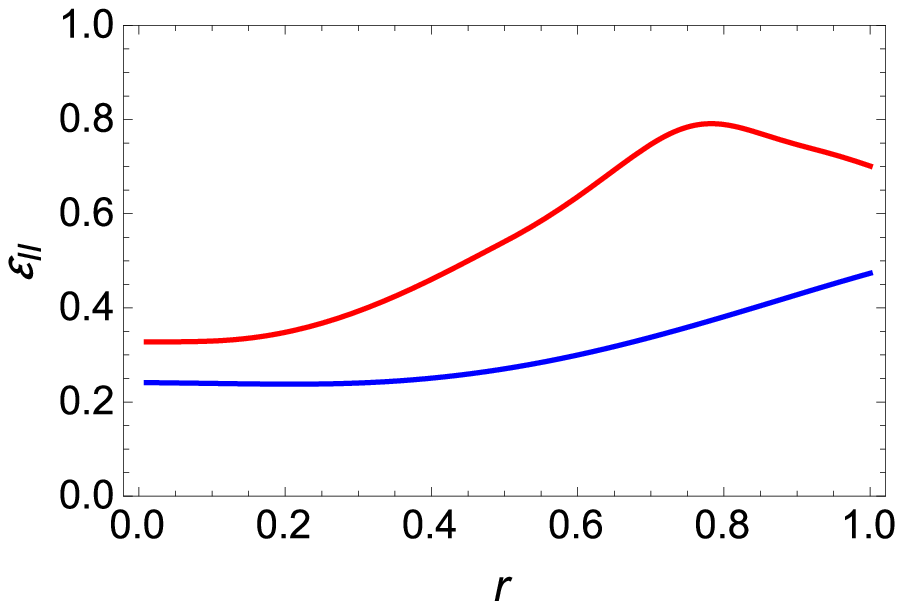} 
\hskip 0.2 truecm  
\includegraphics[scale=0.455]{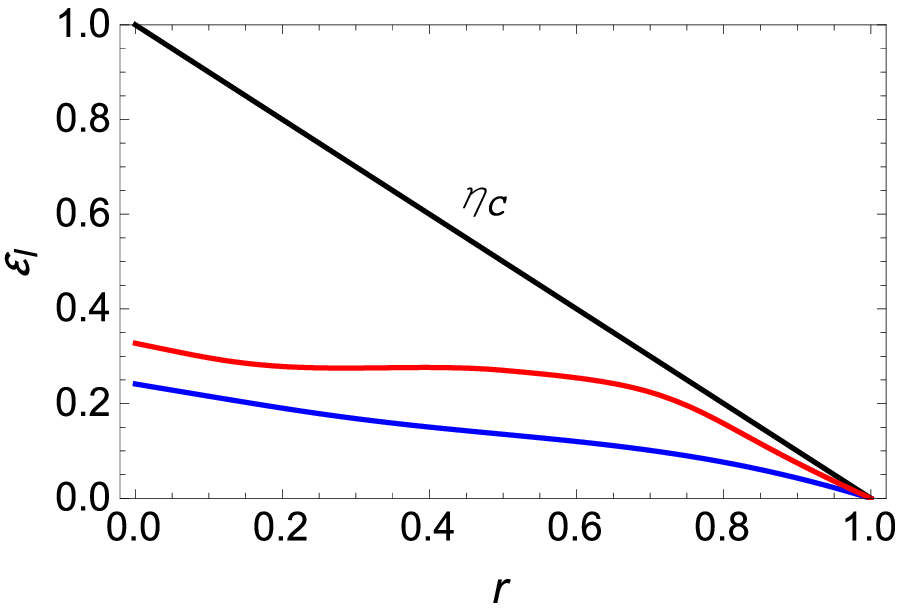}
\end{center}
\caption{(Color online) The efficiencies $\varepsilon_{\rm II}$ (left panel) and $\varepsilon_{\rm I}$ (right panel) 
for $\mu_1=-0.1$ and $\mu_2=-0.5$ (blue line) and  $\mu_1=-0.4$ and $\mu_2=-0.2$ (red line) 
at $\beta_1=2$ and $\tau =0.2$. The black line is the Carnot efficiency.} 
\label{fig3}
\end{figure}

The analytic study of (\ref{s5},\ref{s7}) for generic values of the heat bath parameters 
is rather complicated. Fortunately however, numerics works 
quite well. This fact is illustrated in Fig. \ref{fig3}, where $\varepsilon_{\rm II}$ and $\varepsilon_{\rm I}$ 
are plotted for different values of $\mu_i$. The blue and red lines display the typical behaviour 
for $\mu_2 < \mu_1<0$ and $\mu_1<\mu_2<0$ respectively. 

Finally, we would like to stress that the efficiency $\varepsilon_{\rm II}$ applies to both regimes of 
operation of the quantum junction as a converter of heat to chemical energy or vice versa.

\section{Role of statistics - comparison with fermions} 

In the fermionic case the Pauli exclusion principle 
simplifies the picture. In fact, the $n\geq 2$ emission/absorption 
processes with the same energy are forbidden and the fermionic distribution $\varrho^{\rm f} [j_1]$ 
involves three terms only: 
\begin{equation}
\varrho^{\rm f}[j_1](\xi ) = \sum_{k=-1}^1 p^{\rm f}_k\, \delta(\xi -k\sqrt \tau) \, . 
\label{fe1}
\end{equation} 
The three ``teeth" of the Dirac comb (\ref{fe1}) are 
\begin{equation} 
p^{\rm f}_{\pm 1} = \frac{1}{2}\left (c^{\rm f}_2 \mp c^{\rm f}_1\sqrt {\tau}\right )\, ,\qquad  
p^{\rm f}_0 = 1-c^{\rm f}_2 \, , 
\label{fe2}
\end{equation} 
where 
\begin{equation}
c^{\rm f}_1 \equiv d^{\rm f}_1 - d^{\rm f}_2 \, , \qquad 
c^{\rm f}_2 \equiv d^{\rm f}_1 + d^{\rm f}_2 -2 d^{\rm f}_1 d^{\rm f}_2\, ,  
\label{fe3}
\end{equation}
and 
\begin{equation}
d^{\rm f}_i(\omega ) =  \frac{1}{\e^{\beta_i (\omega -\mu_i)}+1}  
\label{fe4}
\end{equation}  
is the Fermi distribution. It is easily seen that 
\begin{equation}
p^{\rm f}_{-1}+p^{\rm f}_0+p^{\rm f}_{1} = 1\, , \quad p^{\rm f}_{\pm 1}\in [0,1]\, , \quad p^{\rm f} \in [0,1]\, ,  
\label{fep}
\end{equation}
which imply that $\varrho^{\rm f}[j_1]$ is a true probability distribution. 

In terms of the probabilities (\ref{fe2}) the fermionic entropy production distribution takes the form  \cite{MSS-17} 
\begin{equation}
\varrho^{\rm f}[\dS](\sigma ) = \sum_{k=-1}^{1} p^{\rm f}_k\, \delta(\sigma -k\, \gamma_{21}{\sqrt \tau}) \, .  
\label{fe5}
\end{equation} 
The relative moments ($k=1,2,...$) 
\begin{eqnarray}
\M^{\rm f}_{2k-1}[\dS] &=& \tau^{k} \gamma_{21}^{2k-1}\, c^{\rm f}_1\, , 
\label{nmf1}\\
\M^{\rm f}_{2k}[\dS] &=& \tau^{k} \gamma_{21}^{2k}\, c^{\rm f}_2\, , 
\label{nmf2}
\end{eqnarray} 
are much simpler then the bosonic ones (\ref{K}, \ref{me1}) and satisfy the bound 
$\M^{\rm f}_n[\dS] \geq 0$ implementing the second law. 

\begin{figure}[ht]
\begin{center}
\includegraphics[scale=0.455]{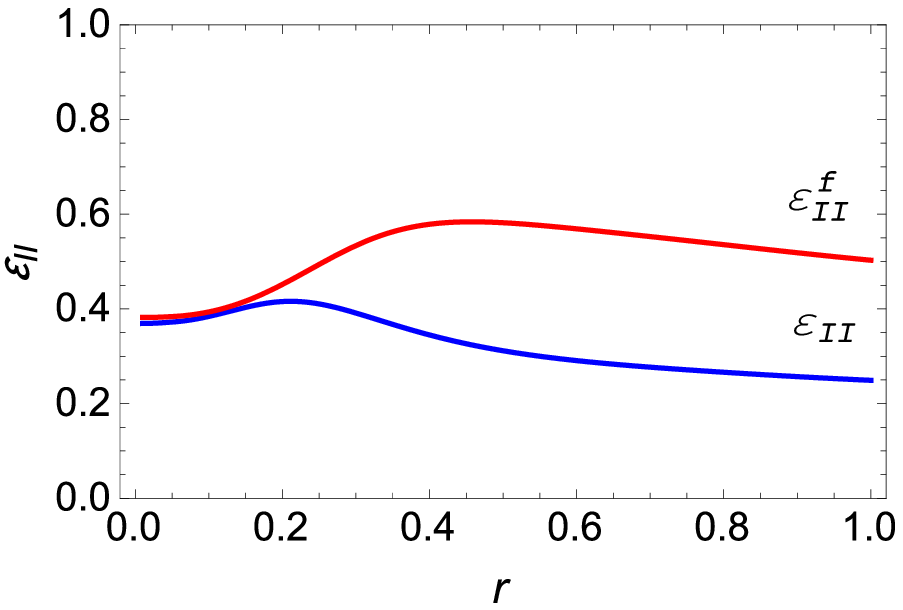} 
\hskip 0.2 truecm  
\includegraphics[scale=0.455]{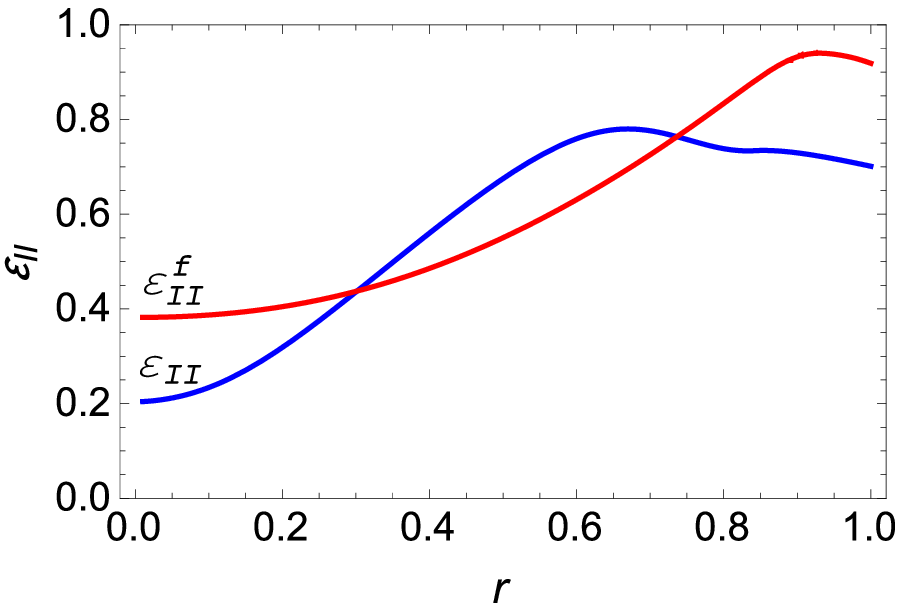}
\end{center}
\caption{(Color online) The fermionic efficiency $\varepsilon^{\rm f}_{\rm II}$ (red line) and the bosonic efficiency $\varepsilon_{\rm II}$ (blue line) 
for $\mu_1= -20$, $\mu_2=-1$ and $\tau=0.2$ at $\beta_1=0.1$ (left panel) and $\beta_1 = 0.01$ (right panel).} 
\label{fig4}
\end{figure}

The difference between the non-equilibrium transport of 
bosons and fermions emerges also by comparing the relative 
efficiencies. From (\ref{fe5}) one infers the entropy productions 
\begin{equation}
\sigma^{\rm f}_\pm (\omega ) = \pm \theta (\gamma_{12}) \gamma_{12} \sqrt{\tau}\, p^{\rm f}_{\mp 1} 
\pm \theta (\gamma_{21}) \gamma_{21} \sqrt{\tau}\, p^{\rm f}_{\pm 1}\, .  
\label{fe6} 
\end{equation} 
Substituting (\ref{fe6}) in (\ref{s3}) one obtains the fermionic versions $\varepsilon_{I}^{\rm f}$ and 
$\varepsilon_{II}^{\rm f}$ of the first and second law efficiencies, which differ from the bosonic ones. 
Fig. \ref{fig4} displays a comparison between $\varepsilon_{II}^{\rm f}$ and its bosonic counterpart 
$\varepsilon_{II}$ at the same heat bath parameters. In the left panel $\varepsilon_{II}^{\rm f}$ exceeds 
$\varepsilon_{II}$. For the same chemical potentials, but at higher temperature $1/\beta_1$ in 
the right panel there is an interval in $r$ for which $\varepsilon_{II} > \varepsilon_{II}^{\rm f}$. 

\section{Discussion}

The main goal of the present paper is to develop a microscopic approach to 
non-equilibrium transport , which takes into account in a systematic way the 
quantum fluctuations at any order. The basic idea is to use the probability distributions 
$\varrho[j_i]$ and $\varrho[\dS]$, generated respectively 
by the $n$-point correlation functions of the particle current and  
entropy production operators for all $n\geq 1$. We have shown that these distributions 
determine a sequence of probabilities $\{p_k(\omega )\, :\, k=0,\pm 1,\pm 2,...\}$ 
associated with the fundamental microscopic processes of emission and 
absorption of particles from the heat reservoirs, driving the system away from 
equilibrium. It turns out that these probabilities fully describe the particle, energy and heat transport. 
Moreover, they determine the quantum entropy production and characterise 
the efficiency at the microscopic level, thus providing universal information  
about the system. 

The above general ideas, which have been illustrated in the paper 
on the example of an exactly solvable model, 
suggest different promising directions for further research. First of all, it will be 
interesting to lift the zero frequency condition and evaluate the distributions 
$\varrho[j_i]$ and $\varrho[\dS]$ in general. In this respect, the study 
\cite{MSS-16} of the second moment $\M[j_i]$ at arbitrary finite frequency 
indicated for instance the relevant impact of bound states on the particle 
transport. This result implies that the recent experimental progress \cite{K-15}-\cite{W-16} in finite 
frequency quantum transport can provide a new valuable tool for bound 
state spectroscopy. 

A further challenging question in the above context concerns the interactions. 
Since our exactly solvable system involves only a boundary interaction at the junction, 
one may wonder about the role of bulk interactions. In this respect the 
analysis of \cite{MS-13} represents a starting point for the study of  
the non-equilibrium Luttinger liquid. Another recently investigated \cite{BPC-18} 
example is the Lieb-Liniger model with contact repulsive interactions. 
The results of \cite{BPC-18} concern the probability distribution $\varrho[\psi^*\psi]$, 
generated by the particle density operator $\psi^*\psi$, and are obtained in a 
specific non-equilibrium regime.  
It is worth mentioning that also in that case $\varrho[\psi^*\psi]$ is a 
Dirac comb distribution, whose coefficients are the counterparts of our 
probabilities (\ref{pr7}). It will be interesting to study the entropy production 
in the Lieb-Liniger case, exploring the influence of the bulk interactions 
on the positivity of the mean entropy production and the higher moments 
of the associated distribution.




\begin{thebibliography}{00} 

\bibitem{BDZ-08} 
I.~Bloch, J.~Dalibard and W.~Zwerger, 
Rev. Mod. Phys. {\bf 80}, 885 (2008). 

\bibitem{CCGOR-11}
M.~A.~Cazalilla, R.~Citro, T.~Giamarchi, E.~Orignac and M.~Rigol,
Rev. Mod. Phys. {\bf 83}, 1405 (2011). 

\bibitem{CS-16}
F.~Chevy and C. Salomon, 
J. Phys. B: At. Mol. Opt. Phys. {\bf 49}, 192001 (2016). 

\bibitem{Z-04}
A. Micheli, A. J. Daley, D. Jaksch and P. Zoller, 
Phys.~Rev.~Lett. {\bf 93}, 140408 (2004). 

\bibitem{Z-07}
J. A. Stickney, D. Z. Anderson and A. A. Zozulya, 
Phys. Rev. A {\bf 75}, 013608 (2007).

\bibitem{P-10}
R. A. Pepino, J. Cooper, D. Meiser, D. Z. Anderson and M. J. Holland,
Phys. Rev. A {\bf 82}, 013640 (2010).

\bibitem{L-57}
R. Landauer, 
IBM J. Res. Dev. {\bf 1}, 233 (1957);  
Philos. Mag. {\bf 21}, 863 (1970).

\bibitem{B-86}
M. B\"uttiker, 
Phys. Rev. Lett. {\bf 57},1761 (1986); 
IBM J. Res. Dev. {\bf 32}, 317 (1988). 

\bibitem{A-80} 
P.~W.~Anderson, D.~J.~Thouless, A.~Abrahams and D.~S.~Fisher, 
Phys.\ Rev.\ B {\bf 22}, 3519 (1980). 

\bibitem{F-81} 
D.~S.~Fisher and P.~A.~Lee, 
Phys.\ Rev.\ B {\bf 23}, 6851 (1981). 

\bibitem{SI-86} 
U.~Sivan and Y.~Imry, 
Phys.\ Rev.\ B {\bf 33}, 551 (1986). 

\bibitem{ML-92} 
Th.~Martin and R.~Landauer, 
Phys.\ Rev.\ B {\bf 45}, 1742 (1992). 

\bibitem{B-92}
M. B\"uttiker, 
Phys.\ Rev.\ B  {\bf 46}, 12485 (1992). 

\bibitem{JB-96} 
M.~J.~M.~de Jong and C.~W.~J.~Beenakker, {\it Shot noise in mesoscopic systems}, 
in {\it Mesoscopic Electron Transport}, 
edited by L.~L.~Sohn, L.~P.~Kouwenhoven and G.~Schoen, 
NATO ASI Series 345 (Kluwer Academic Publishers, Dordrecht, 1997), 225.

\bibitem{BB-00}
Ya.~M.~Blanter and M. B\"uttiker, 
Phys.\ Rep.\  {\bf 336}, 1 (2000). 

\bibitem{MSS-15}  
M.~Mintchev, L.~Santoni and P.~Sorba, 
J.\ Phys.\ A {\bf 48}, 285002 (2015). 

\bibitem{MSS-17an}  
M.~Mintchev, L.~Santoni and P.~Sorba, 
Annalen~der~Physik {\bf 529}, 1600274 (2017). 

\bibitem{K-87} V.~K.~Khlus, Sov. Phys. JETP {\bf 66}, 1243 (1987). 

\bibitem{L-89} 
G.~B.~Lesovik, 
JETP~Lett. {\bf 49}, 592 (1989).

\bibitem{LL-92}
L.~S.~Levitov and G.~B.~Lesovik, 
JETP~Lett. {\bf 55}, 555 (1992). 

\bibitem{LLL-96} 
L.~S.~Levitov, H.~Lee and G.~B.~Lesovik, 
J. Math. Phys. {\bf 37}, 4845 (1996). 

\bibitem{LC-03}
G.~B.~Lesovik and N.~M.~Chtchelkatchev, 
JETP~Lett. {\bf 77}, 393 (2003). 

\bibitem{K-03} 
I.~Klich, 
{\it An elementary derivation of Levitov's formula}, 
in {\it Quantum Noise in Mesoscopic Physics}, 
edited by Y.~V.~Nazarov, (Kluwer Academic Publishers, Dordrecht 2003), 397.

\bibitem{GGM-03} 
D.~B.~Gutman, Y.~Gefen and A.~D.~Mirlin, 
{\it High cumulants of current fluctuations out of equilibrium}, 
in {\it Quantum Noise in Mesoscopic Physics}, 
edited by Y.~V.~Nazarov, (Kluwer Academic Publishers, Dordrecht 2003), 497.

\bibitem{ELB-09}
M.~Esposito,  U.~Harbola and S.~Mukamel, 
Rev. Mod. Phys. {\bf 81}, 1665 (2009). 

\bibitem{Gasp-15}  
P.~Gaspard, 
Annalen~der~Physik {\bf 527}, 663 (2015). 

\bibitem{MSS-16}  
M.~Mintchev, L.~Santoni and P.~Sorba, 
J.\ Phys.\ A {\bf 49},  265002 (2016). 

\bibitem{M-11} 
M.~Mintchev, 
J.\ Phys.\ A {\bf 44}, 415201 (2011). 

\bibitem{MSS-14}  
M.~Mintchev, L.~Santoni and P.~Sorba, 
J.\ Phys.\ A {\bf 48}, 055003 (2015). 

\bibitem{MSS-17} 
M.~Mintchev, L.~Santoni and P.~Sorba, 
Phys.\ Rev.\ E {\bf 96}, 052124 (2017). 

\bibitem{Bejan}
A. Bejan, {\it Advanced Engineering Thermodynamics}, 
(John Wiley and Sons, 2016). 

\bibitem{Callen}
H. B. Callen, {\it Thermodynamics and an Introduction to Thermostatistics}, 
(John Wiley and Sons, 1960). 

\bibitem{JP-01} 
V.~Jaksi$\check{\rm c}$ and C.-A.~Pillet, 
Comm. Math. Phys. {\bf 217}, 285 (2001). 

\bibitem{N-07} 
G.~Nenciu, 
J. Math. Phys. {\bf 48}, 033302 (2007).  

\bibitem{ks-00}
V.~Kostrykin and R.~Schrader, 
Fortschr. Phys. {\bf 48}, 703 (2000). 

\bibitem{h-00}
M.~Harmer, 
J.\ Phys.\ A {\bf 33}, 9015 (2000). 

\bibitem{ST-70} 
J.~A.~Shohat and J.~D.~Tamarkin, {\it The problem of moments} 
(American Mathematical Society, Providence, Rhode Island 1970). 

\bibitem{Glauber}
R.~J.~Glauber, 
{\it Quantum Theory of Optical Coherence}, 
(WILEY-VCH Verlag GmbH and Co. KGaA, Weinheim 2007). 

\bibitem{BHNOW-15}
F.~Brandao, M.~Horodecki N.~Ng, J.~Oppenheim and S.~Wehner, 
Proc. Natl. Acad. Sci. U.S.A. {\bf 112}, 3275 (2015). 

\bibitem{CSHO-15}
P.~Swiklinski, M.~studzinski, M.~Horodecki and J.~Oppenheim, 
Phys. Rev. Lett. {\bf 115}, 210403 (2015). 


\bibitem{K-15}
S. Kolkowitz et al, Science {\bf 347}, 1129 (2015). 

\bibitem{T-16}
E. S. Tikhonov et al, Nature Sci. Rep. {\bf 6}, 30621 (2016). 

\bibitem{W-16}
Q. Weng et al, {\it Imagining non-local electron transport via local excess
noise}, arXiv: 1610.01711. 

\bibitem{MS-13}  
M.~Mintchev and P.~Sorba, 
J.\ Phys.\ A {\bf 46},  095006 (2013). 

\bibitem{BPC-18}
A.~Bastianello, L.~Piroli and P.~Calabrese, 
{\it Exact local correlations and full counting statistics for arbitrary states of the 
one-dimensional interacting Bose gas}, arXiv::1802:02115. 


\end{thebibliography}
\end{document}